\documentstyle[11pt]{article}
\begin{document}
\begin{flushright} 
{hep-th/0210304}
\end{flushright}
\begin{center}
{\large \bf From Wigner--In{\"o}n{\"u} Group Contraction 
to Contractions of Algebraic Structures}  \\
{N.A. Gromov}                               \\
{Department of Mathematics, Institute of Mathematics and Mechanics,\\
Ural Division, Russian Academy of Sciences,
Syktyvkar 167982, Russia \\
E-mail: gromov@dm.komisc.ru}
 \end{center}

\abstract{
The development of the notion of group contraction first introduced
by E. In{\"o}n{\"u} and E.P. Wigner in 1953 is briefly reviewed.
The fundamental role of the idea of degenerate transformations is stressed.
The significance of contractions of algebraic structures for exactly
solvable problems of mathematical physics is noticed.}

\section{ Introduction}

Starting with the fact that classical mechanics is
a limiting case of relativistic mechanics and therefore
the Galilei group must be in some sense a limiting case
of the relativistic mechanics' group E.P. Wigner and
E. In{\"o}n{\"u} in 1953 have introduced \cite{IW}
 a new notion in group theory, namely,
a {\it contraction} of group (Lie algebra and their
representations) as follows: "We shall call the operation
of obtaining a new group by a singular transformation of the
infinitesimal elements of the old group a {\it contraction}
of the latter." This concept is based on the principal idea:
 only singular transformation of Lie algebra gives in result
a new nonisomorphic algebra.

The concept of contraction turned out to be fruitful and
was formulated \cite{G-95} as the contraction
of an arbitrary algebraic structure $(M,*)$
with the help of $\epsilon$-depending map
$$
 \phi_{\epsilon} : (M,*) \rightarrow (N,*'),
$$
where $(N,*')$ is the algebraic structure of the same type,
isomorphic to $(M,*)$ for  $ \epsilon \neq 0 $ and nonisomorphic to
the initial one for $ \epsilon = 0. $
Except for original Wigner--In{\"o}n{\"u} group (algebra) 
contractions the contractions of Lie superalgebra was introduced \cite{RT}
 and after that more generals  graded contractions \cite{P}, \cite{MP}
  which preserve an arbitrary grading of Lie algebra
and may be applied to an infinitedimensional algebras \cite{KGK}.
 Under contractions of bialgebra \cite{VG}
  Lie algebra structure and cocom\-mutator are conserved.
 Hopf algebra ( quantum algebra \cite{C1}, \cite{C2} 
and quantum group  \cite{G})
contractions are introduced in such a way, that in the limit 
$ \epsilon \rightarrow 0 $
a new expressions for coproduct, counit and antipode
are consistent with the Hopf algebra axioms.
Recently contractions of the algebraic structures with bilinear 
products of arbitrary nature on sections of finite-dimensional 
vector bundles was presented  \cite{CGM}.

\section{ Contractions of algebraic structures}

 A {\bf Lie algebra} $ L $ is a vector space $ V $ equipped
with a Lie bracket $ [ \ , \ ]: L \otimes L \rightarrow L, $
which is linear in both arguments, antisymmetric and satisfy the
Jacobi identity. So the structure of Lie algebra is as follows:
$ L=(V,[ \ , \ ])$. It is well known fact that the non--degenerate
linear transformations $ \Phi: V \rightarrow V, \; \det{\Phi} \not = 0 $
of vector space $ V $ give in result the Lie algebra $ L^{'} $
which is  isomorphic to the initial one's $ L $. How to receive a new Lie
algebra from $ L $? The answer was given fifty years ago 
by E. In{\"o}n{\"u} and E.P. Wigner  \cite{IW}: 
it is necessary to use a degenerate transformation
of $ V $. Let us regard $ \Phi_{\epsilon}: V \rightarrow V $ such that
$ \det \Phi_{\epsilon} \not =0 $ for $ \epsilon \not =0 $, but
$ \det \Phi_{\epsilon}=0 $ for $ \epsilon=0 $. Then if the limit
\begin{equation}
[ \ , \ ]^{'}=\lim_{\epsilon \to 0}\Phi_{\epsilon}\circ [ \ , \ ]\circ
 (\Phi_{\epsilon}^{-1} \otimes \Phi_{\epsilon}^{-1})
\label{1}
\end{equation}
exist, we have the new Lie algebra $ L^{'}=(V,[ \ , \ ]^{'}) $
with the same underlying vector space. The limit (\ref{1}) is
called a {\it contraction of Lie algebra}.

A {\bf Lie superalgebra} has the structure
$L_s=(V=V_0 \oplus V_1,< \ , \ >,Z_2)=L^0_s \oplus L^1_s,$
where $< \ , \ >$ is a bilinear product (commutator or anticommutator),
$Z_2$ is a grading group, such that $V_0$ and $V_1$ are even and odd 
parts of vector space $V,$ respectively, $L_s^0$ and $L_s^1$ are
even and odd parts of superalgebra $L_s.$ The nondegenerate limit \cite{RT}
\begin{equation}
< \ , \ >^{'}=\lim_{\epsilon \to 0}\Phi_{\epsilon}\circ < \ , \ >\circ
 (\Phi_{\epsilon}^{-1} \otimes \Phi_{\epsilon}^{-1}),
\label{1-1}
\end{equation}
where $\Phi_{\epsilon} $ preserves $Z_2-$grading, gives in result
contracted superalgebra  
$L_s^{'}=(V=V_0 \oplus V_1,< \ , \ >^{'},Z_2)=L^{0'}_s \oplus L^{1'}_s.$

{\bf Graded contractions} were introduced in \cite{P}, \cite{MP} and
deal with the structure $ (L,\Gamma) $, where $ L $ is a Lie
algebra and $ \Gamma $ is a finite Abelian group. A grading
of $ L $ by $ \Gamma $ means the decomposition of the vector space
$ V $ of $ L $ into a direct sum of subspaces
\begin{equation}
V=\bigoplus_{\mu \in \Gamma} V_{\mu}
\label{2}
\end{equation}
such that for every choice of $  X \in V_{\mu}, \  Y \in V_{\nu} $,
if $ [X,Y] \not =0 $, then $ [X,Y] \in V_{\mu+\nu} $.
Symbolically
\begin{equation}
0 \not = [V_{\mu},V_{\nu}] \subseteq V_{\mu+ \nu}, \  \mu, \nu, \mu+\nu \in
 \Gamma.
\label{3}
\end{equation}
In this case the map $ \Phi_{\epsilon} $ additionally to
existence of (\ref{1}) has to preserve the grading equations
(\ref{2}), (\ref{3}). Graded contractions leads to a structure
$ (L^{'}, \Gamma)=(V,[ \ , \ ]^{'}, \Gamma) $ with the same
underlying $ V $ and the same grading group $ \Gamma $.
Graded contractions of the infinitedimensional Virasoro algebras 
were regarded in \cite{KGK}.

 A {\bf Lie bialgebra} $ (L,\eta) $ is a Lie algebra $ L $ endowed with
a cocomutator  $ \eta:L \rightarrow L \otimes L $ such that

i) $ \eta $ is 1--cocycle, i.e.
\begin{equation}
\eta([X,Y])=[\eta(X), 1 \otimes Y+Y \otimes 1]+
[1 \otimes X+X \otimes 1, \eta(Y)], \ \forall X,Y \in L,
\label{4}
\end{equation}

ii) the dual map $ \eta^{*}:L^{*} \otimes L^{*} \rightarrow L^{*} $
is a Lie bracket on $ L^{*} $.

Contracted Lie brackets are obtained by (\ref{1}).
Contraction of cocommutator $ \eta $ may be realized by the mapping
$ \Psi_{\epsilon}:L \rightarrow L^{'} $, which is in general
different from $ \Phi_{\epsilon} $, as follows:
\begin{equation}
\eta^{'}=\lim_{\epsilon \to 0}(\Psi_{\epsilon} \otimes \Psi_{\epsilon})\circ
 {\eta}\circ {\Psi}-{\epsilon}^{-1}.
\label{5}
\end{equation}
The consistency of $ [ \ , \ ] $ and $ \eta $
 gives the another expression for cocommutator
\begin{equation}
\eta^{'}=\lim_{\epsilon \to 0}(\Psi_{\epsilon} \otimes \Psi_{\epsilon})\circ
 {\eta}\circ {\Phi}^{-1}_{\epsilon}.
\label{6}
\end{equation}
If all limits (\ref{1}), (\ref{5}), (\ref{6}) exist, then
$ (L^{'},{\eta}^{'}) $ is {\it a contracted Lie bialgebra} \cite{VG}.

An associative algebra $ A $ is said to be a {\bf Hopf algebra} if there
exist maps: coproduct $ \Delta: A \rightarrow A \otimes A, $
counit $ u: A \rightarrow {\rm C} $ and antipode
$ \gamma: A \rightarrow A $, such that $ \forall a \in A $
 the following axiomas are hold
\begin{eqnarray}
H_1:  &  (id  \otimes   \Delta)\Delta(a)=(   \Delta   \otimes
 id)\Delta(a), \nonumber \\
H_2: &(id \otimes u)\Delta(a)=(u \otimes id)\triangle(a), \\
H_3: & m((id \otimes \gamma)\Delta(a))=m((\gamma \otimes id)\Delta(a)),
 \nonumber
\label{7}
\end{eqnarray}
where $ m(a \otimes b)=ab $ is the usual multiplication.
So the Hopf algebra structure is $ H=(A,\Delta,u,\epsilon)$.
Contractions of coproduct, counit and antipode
are realized by the same mapping $\chi_{\epsilon}$ of $ A $. 
 Therefore, if the limits
\begin{equation}
\Delta^{'}=\lim_{\epsilon \to 0}({\chi}_{\epsilon} \otimes
 {\chi}_{\epsilon})\circ {\Delta}\circ {\chi}_{\epsilon}^{-1}, \quad
u^{'}=\lim_{\epsilon \to 0}u\circ {\chi}_{\epsilon}^{-1}, \quad
\gamma^{'}=\lim_{\epsilon \to 0}{\chi}_{\epsilon}\circ {\gamma}\circ
 {\chi}_{\epsilon}^{-1}.
\label{8}
\end{equation}
exist, then $ H^{'}=(A,{\Delta}^{'},u^{'},{\epsilon}^{'}) $
is {\it a contracted Hopf algebra.}
Contractions of Hopf algebras on the level of quantum  algebras
 were first developed in \cite{C1},\cite{C2} and on the level of
quantum groups in \cite{G}.

The general conception of contraction of algebraic structures
was recently developed \cite{CGM} to contractions of  $(A,*),$ where
{\bf $A$ is a set of smooth sections of vector bundle $E$ over 
a manifold $ M,\; *: A \times A \rightarrow A $ is a point-wise
continuous bilinear product }
are given by the following 

{\bf Theorem.} {\it Let $N: E  \rightarrow E $ be a smooth 
$(1,1)-$tensor. Denote by $U(\epsilon)=\epsilon I+N$ a deformation
of $N,$ by $E=E^1 \oplus E^2$ the Riesz decomposition of $E$ 
relative to $N,$ and by $A^1$ the set of smooth sections of $E^1.$
Then the limit
\begin{equation}
X *_N Y = \lim_{\epsilon \to 0}U^{-1}(\epsilon)
 (U(\epsilon)(X) * U(\epsilon)(Y))
\label{9}
\end{equation}
exists for all $X,Y\in A$ and defines a new (contracted) bilinear
product on $A$ if and only if the torsion
\begin{equation}
T_N(X * Y) = N(X)*N(Y) - N(N(X)*Y +X*N(Y)) + N^2(X*Y)
\label{10}
\end{equation}
takes values in $N(A^1).$}

Lie algebras, Lie algebroids and Poisson brackets were given 
as an examples of the developed theory. In particular, 
Wigner--In{\"o}n{\"u} Lie algebra contractions correspond to
the vector bundle $E$ over a single point with 
$ A=E,\; *=[ \ , \ ],\; U(\epsilon)=P + \epsilon (I-P),$ where $P$ is
a projection.

\section{Conclusions}\label{concl}

The notion of contraction first introduced by E. In{\"o}n{\"u}
and E.P. Wigner for Lie groups and algebras is now extend
on new algebraic structures. Some of them are mention in this
talk. But the fundamental idea of degenerate transformation
is presented in all cases.

It is well known that often a symmetry is a necessary condition 
for explicit solutions of many mathematical and physical problems.
And at the second half of the last century a models 
of the mathematical physics are usually formulated with the help of
different mathematical structures (Lie groups and algebras, 
superalgebras, quantum groups etc.), which are a mathematical tools 
for describing symmetries. Above-mentioned algebraic structures 
constitute the base of exactly solvable problems of mathematical physics
and its contractions are correspond to different limiting cases of these
problems.

\end{document}